\documentclass[12pt]{article}
\usepackage{bbm}
\usepackage{latexsym}
\begin{document}
\title{\normalsize \hfill UWThPh-2003-31 \\[1cm] \LARGE
Models of maximal atmospheric neutrino mixing\footnote{Talk presented
by W.\ Grimus at the XXVII International Conference of Theoretical Physics, 
\textit{Matter to the Deepest: Recent Developments in
Physics of Fundamental Interactions},
Ustro\'n, Poland, 15--21 September 2003}}
\author{\setcounter{footnote}{3}
Walter Grimus\thanks{E-mail: walter.grimus@univie.ac.at} \\
\small Institut f\"ur Theoretische Physik, Universit\"at Wien \\
\small Boltzmanngasse 5, A--1090 Wien, Austria \\*[3.6mm]
Lu\'{\i}s Lavoura\thanks{E-mail: balio@cfif.ist.utl.pt} \\
\small Universidade T\'ecnica de Lisboa \\
\small Centro de F\'\i sica das Interac\c c\~oes Fundamentais \\
\small Instituto Superior T\'ecnico, P--1049-001 Lisboa, Portugal \\*[4.6mm] }

\date{3 October 2003}

\maketitle

\begin{abstract}
We discuss two types of neutrino mass matrices which both give
$\theta_{23} = 45^\circ$, i.e., a maximal atmospheric mixing angle. We
review three models, based on the seesaw mechanism and on simple
extensions of the scalar sector of the Standard Model, where those mass
matrices are obtained from symmetries. 
\\[3mm]

\noindent
PACS numbers: 16.60.Pq, 14.60.St
\end{abstract}

\newpage

\section{Motivation and introduction}

In the last years great progress has been made in the measurements and
understanding of the solar and atmospheric neutrino fluxes, 
and the oscillation solutions \cite{pontecorvo} for the  
solar and atmospheric neutrino deficits have been
established---for recent reviews see, for instance, Ref.~\cite{reviews}.
At the same time, the amazing precision of the Solar Standard
Model \cite{bahcall}, a necessary ingredient for the evaluation of the
solar neutrino data, has also become evident. 
The measurement of the total active $^8$B neutrino flux with
enhanced neutral-current sensitivity (salt phase) by the SNO
Collaboration \cite{SNO} has further 
corroborated this picture for the solar neutrinos, and it
has considerably reduced the allowed region in the solar oscillation
parameters---for analyses including the SNO result 
see Refs.~\cite{balantekin,fogli,maltoni,aliani,goswami,smirnov}.
The present knowledge of the neutrino mixing angles can be
summarized in the following way.
For the atmospheric mixing angle
the Super-Kamiokande Collaboration \cite{SK} has obtained the bound
$\sin^2 2\theta_\mathrm{atm} > 0.9$ at 90\% CL, which
corresponds to $\theta_\mathrm{atm} = 45^\circ \pm 9^\circ$. 
The allowed range for the solar mixing angle can be read off from the
regions allowed at 90\% CL in the above-mentioned
papers, and is estimated as 
$\theta_\odot \sim 33^\circ \!\!
\renewcommand{\arraystretch}{0.6}
\begin{array}{c} \scriptstyle +4^\circ \\ \scriptstyle -3^\circ 
\end{array}$.
Furthermore, from the new SNO data it follows that  
$\theta_\odot \neq 45^\circ$ at the $5\, \sigma$ level \cite{SNO}.
A further interesting development is that solar neutrino data have
become numerically important
in a three-neutrino analysis of the mixing angle
$\theta_{13}$ \cite{maltoni,goswami}; at the $3\, \sigma$ level 
Ref.~\cite{maltoni} has obtained 
$\sin^2 \theta_{13} < 0.044$.
In addition, it has been established that, in the oscillation solution 
for the solar deficit, matter effects \cite{MSW} play a decisive role
\cite{fogli}. 

In the course of time, the best-fit value of the atmospheric mixing
has always remained stable at $45^\circ$. Therefore, 
we believe that the above results provide a motivation to 
search for models which have \textit{maximal atmospheric neutrino mixing 
enforced by a symmetry}  
and \textit{large but non-maximal solar neutrino mixing}.

In the following, we will only discuss models of massive Majorana
neutrinos with a mass term 
\begin{equation}\label{mass}
\mathcal{L}_\mathrm{mass}
= \frac{1}{2}\, \nu_L^T C^{-1} \mathcal{M}_\nu \nu_L + \mathrm{H.c.}
\end{equation}
We consider the following two mass matrices: 
\begin{eqnarray}
\mathrm{M}1: && 
\mathcal{M}_\nu = \left( \begin{array}{ccc} x & y & y \\ y & z & w \\
y & w & z \end{array} \right)
\quad \mathrm{with} \quad 
x, \, y, \, z, \, w \in \mathbbm{C} \,,
\label{M1} \\
\mathrm{M}2: &&
\mathcal{M}_\nu = \left( \begin{array}{ccc} a & r & r^\ast \\ r & s & b \\
r^\ast & b & s^\ast \end{array} \right) 
\quad \mathrm{with} \quad  
r, \, s \in \mathbbm{C} \,, \, 
a, \, b \in \mathbbm{R} \,.
\label{M2}
\end{eqnarray}
These mass matrices are defined 
in the basis where the charged-lepton mass matrix is diagonal. 
Phenomenological discussions of the matrix M1 can be found in many
papers---see, e.g., Ref.~\cite{lam}---whereas M2
was recently found by Babu, Ma, and Valle
in the context of models based on the group $A_4$ \cite{ma1,ma2}. 
We will see in the
following that M1 and M2 have maximal atmospheric neutrino mixing.
For
an attempt to obtain $\theta_\mathrm{atm} = 45^\circ$ based on the
group $S_3$ see Ref.~\cite{kubo}.

The subject of the talk is the following: 
\begin{itemize}
\item[$\star$]
Discussion of the phenomenology of M1, M2;
\item[$\star$]
Review of the models of Refs.~\cite{GL01,GL03} and of
Ref.~\cite{GLcp}, which produce M1 and M2, respectively, by symmetries.
\end{itemize}

\section{Phenomenology of the mass matrices M1, M2}

The matrices M1, M2 can be algebraically characterized in a very
simple way. Defining a unitary matrix 
\begin{equation}\label{S}
S =
\left( \begin{array}{ccc} 1 & 0 & 0 \\ 0 & 0 & 1 \\ 0 & 1 & 0
\end{array} \right),
\end{equation}
the relations 
\begin{eqnarray}
M1: && S \mathcal{M}_\nu S = \mathcal{M}_\nu \,,
\label{M1def} \\
M2: && S \mathcal{M}_\nu S = \mathcal{M}_\nu^* 
\label{M2def}
\end{eqnarray}
can be conceived as defining M1 and M2, respectively. 

The Majorana mass matrix $\mathcal{M}_\nu$ is diagonalized by 
\begin{equation}\label{V}
V^T \mathcal{M}_\nu V = \mathrm{diag}\, ( m_1, m_2, m_3 ) \,,
\end{equation}
where the real and non-negative neutrino masses
have been denoted by $m_j$ ($j=1,2,3$). 
The matrix $V$ is decomposed as 
\begin{equation}\label{Vdecomp}
V = 
e^{i\hat \alpha}
U_{23} U_{13} U_{12}\,
{\rm diag} \left( 1, e^{i \beta_1}, e^{i \beta_2} \right).
\end{equation}
The diagonal phase matrix 
$e^{i\hat \alpha} = \mathrm{diag}\,
\left( e^{i \alpha_1}, e^{i \alpha_2}, e^{i \alpha_3} \right)$
contains unphysical phases which can be absorbed into the
charged-lepton fields. The unitary matrices $U_{23}$, $U_{13}$,
$U_{12}$ are given by  
\begin{eqnarray}
U_{23} &=& \left( \begin{array}{ccc} 1 & 0 & 0 \\
0 & c_{23} & s_{23} \\ 
0 & - s_{23} & c_{23} \end{array} \right), \\
U_{13} &=& 
\left( \begin{array}{ccc} c_{13} & 0 & s_{13} e^{-i \delta} \\
0 & 1 & 0 \\ - s_{13} e^{i\delta} & 0 & c_{13} \end{array} \right), \\
U_{12} &=&
\left( \begin{array}{ccc} c_{12} & s_{12} & 0 \\
- s_{12} & c_{12} & 0 \\ 0 & 0 & 1 \end{array} \right),
\end{eqnarray}
respectively. Here, the notation $c_{12} \equiv \cos \theta_{12}$, 
etc. is used. The phases 
$\beta_1$, $\beta_2$ are the so-called
Majorana phases (only $2 \beta_1$ and $2 \beta_2$ are physical).
The neutrino mixing matrix $U = U_{23} U_{13} U_{12}$
contains the
$CP$-violating phase $\delta$, which is, in principle, accessible in
neutrino oscillations. Our convention for the mixing matrix $U$ is the
same as the convention for the CKM matrix used
in the Review of Particle Properties \cite{RPP} (RPP convention).
Note that $\theta_{12} \equiv \theta_\odot$ and 
$\theta_{23} \equiv \theta_\mathrm{atm}$. 

\subsection{Phenomenology of M1}

Starting with the evident eigenvector relation
\begin{equation}\label{eigen}
\left( \begin{array}{ccc} x & y & y \\ y & z & w \\
y & w & z \end{array} \right) 
\left( \begin{array}{r} 0 \\ 1 \\ -1 \end{array} \right) =
(z - w) 
\left( \begin{array}{r} 0 \\ 1 \\ -1 \end{array} \right),
\end{equation}
it is easy to check that the mixing matrix in the RPP convention
is given by 
\begin{equation}\label{UM1}
U = \left( \begin{array}{ccc}
\cos \theta &  \sin \theta & 0 \\
-\sin \theta/\sqrt{2} & \cos \theta/\sqrt{2} 
& 1/\sqrt{2} \\
\sin \theta/\sqrt{2} & -\cos \theta/\sqrt{2} 
& 1/\sqrt{2}
\end{array} \right). 
\end{equation}
Thus we obtain the following results for the neutrino mixing angles:
\begin{equation}\label{M1result}
\mathrm{M}1: \quad \theta_{13} =  0^\circ \,, \quad 
\theta_{23} = 45^\circ \,, \quad
\theta_{12} \equiv \theta \; \mbox{arbitrary.}
\end{equation}
Furthermore, Eq.~(\ref{eigen}) gives
$m_3 = | z - w |$. The neutrino masses in the case of the mass matrix
M1 are free, i.e., no relations among themselves or with the mixing
angles are obtained. The parameter 
$\sin^2 2\theta_\mathrm{atm} = 
4\, |U_{\mu 3}|^2 \left( 1 - |U_{\mu 3}|^2 \right)$, which is probed
in atmospheric and long-baseline experiments,
is exactly equal to 1.  

On the other hand, if one wishes to use the form (\ref{UM1}) of the
mixing matrix $U$ as input and work back to $\mathcal{M}_\nu$, it is
easy to see that a mass matrix of the form M1 ensues \cite{lam}. 

\subsection{Phenomenology of M2}

With relation (\ref{M2def}) and the physical requirement of a 
non-degenerate three-neutrino mass spectrum one can show that the matrix
$V$ of Eq.~(\ref{V}) must fulfill the condition \cite{GLcp}
\begin{equation}\label{Vcond}
SV^* = V X \,,
\end{equation}
where $X$ is diagonal phase matrix. From this equation it follows
immediately that 
\begin{equation}\label{mu=tau}
|U_{\mu j}| = |U_{\tau j}|\;\: \forall j = 1,2,3 \,.
\end{equation}
Equation~(\ref{mu=tau}) was originally proposed by 
Harrison and Scott \cite{scott}.

Before we proceed further, we note that the sets of matrices of type M1 and
M2 have a non-vanishing overlap; e.g., if a matrix of type M2 is real,
then it is automatically of type M1 also. 
It has been shown in Ref.~\cite{GLcp} that for matrices of type M2 one
has 
\begin{equation}\label{theta13nonzero}
\sin \theta_{13} = 0 \Leftrightarrow  r^2 s^* \in \mathbbm{R} \,.
\end{equation}
One direction of this equivalence is easy to demonstrate, namely if 
$r^2 s^* \in \mathbbm{R}$ then by rephasing one obtains
a matrix of type M1 from a matrix of type M2.
Thus, in the following we will always assume that 
$r^2 s^* \not\in \mathbbm{R}$ for matrices of type M2, in order to
genuinely distinguish them from matrices of type M1.

Then for the matrix M2 one has the following results \cite{ma1,ma2,GLcp}:
\begin{equation}\label{M2result}
\theta_{23} = 45^\circ, \; e^{i\delta} = \pm i, \;
  e^{i\beta_{1,2}} = 1\: \mbox{or}\: i \,.
\end{equation}
The first two results follow readily from relation (\ref{mu=tau}) and
the parameterization 
\begin{equation}
U = 
\left( \begin{array}{ccc}
c_{12} c_{13} & s_{12} c_{13} & s_{13} e^{-i\delta} \\
- s_{12} c_{23} - c_{12} s_{23} s_{13} e^{i\delta} &
c_{12} c_{23} - s_{12} s_{23} s_{13} e^{i\delta} &
s_{23} c_{13} \\
s_{12} s_{23} - c_{12} c_{23} s_{13} e^{i\delta} &
- c_{12} s_{23} - s_{12} c_{23} s_{13} e^{i\delta} &
c_{23} c_{13} \end{array} \right)
\end{equation}
of the mixing matrix. Furthermore, we now have 
$\sin^2 2\theta_\mathrm{atm} = 
4\, |U_{\mu 3}|^2 \left( 1 - |U_{\mu 3}|^2 \right) = 1 - s_{13}^4$; 
for practical purposes this quantity is equal to 1,
due to the smallness of $s_{13}^4$.

\section{The seesaw mechanism with soft breaking of the family lepton numbers}

Now we consider the lepton sector of the Standard Model (SM) with an
arbitrary number $n_H$ of Higgs doublets $\phi_j$,
supplemented by three right-handed neutrino singlets $\nu_R$, 
and allow for lepton number violation. Thus we consider the
Lagrangian 
\begin{eqnarray}
\mathcal{L} & = & \cdots
- \left[ \sum_j \left(
\bar \ell_R \phi_j^\dagger \Gamma_j + 
\bar \nu_R {\tilde\phi_j}^\dagger \Delta_j \right) D_L
+ \mbox{H.c.} \right] \nonumber \\ &&
+ \left( \frac{1}{2}\, \nu_R^T C^{-1} \! M_R^* \nu_R + \mbox{H.c.} \right).
\label{Lsee}
\end{eqnarray}
The charged-lepton singlets are denoted by $\ell_R$ and the lepton
doublets by $D_L$. The dots indicate the gauge part of $\mathcal{L}$.
The mass matrix $M_R$ of the right-handed neutrino singlets is
symmetric. The mass matrix of the charged leptons and the so-called
Dirac mass matrix in the neutrino sector are given by 
\begin{equation}\label{massterms}
M_\ell = \frac{1}{\sqrt{2}} \sum_j v_j^\ast \Gamma_j \,,
\quad 
M_D = \frac{1}{\sqrt{2}} \sum_j v_j \Delta_j \,,
\end{equation}
respectively, with the vacuum expectation values (VEVs) 
$\left\langle \phi_j^0 \right\rangle_0 = v_j/\sqrt{2}$.
The total Majorana mass matrix for left-handed neutrino
fields is obtained as 
\begin{equation}
\setlength{\arraycolsep}{2pt}
\mathcal{M}_{D+M} = \left( \begin{array}{cc}
0 & M_D^T \\ M_D & M_R 	   \end{array} \right)
\quad \mbox{for} \quad
\left( \begin{array}{c} \nu_L \\ C (\bar\nu_R)^T \end{array} \right).
\end{equation}
With the assumption  $m_D \ll m_R$, where 
$m_D$ and $m_R$ are the scales of $M_D$ and $M_R$, respectively, 
the seesaw mechanism \cite{seesaw} is obtained where 
\begin{equation}\label{Mseesaw}
\mathcal{M}_\nu = -M_D^T M_R^{-1} M_D
\end{equation}
for the three light neutrinos. 

Diagonalization of $M_\ell$ proceeds via 
$(U_R^\ell)^\dagger M_\ell U_L^\ell = \hat m_\ell$ with two unitary
matrices $U_{R,L}^\ell$. Then the neutrino mixing matrix is given by  
$U_M = (U_L^\ell)^\dagger V$, where $V$ is defined in
Eq.~(\ref{V}). Thus with the seesaw mechanism there are three 
sources for neutrino mixing: $M_\ell$, $M_D$, and $M_R$. 
We may choose as a typical neutrino mass
$m_\nu \sim \sqrt{\Delta m^2_\mathrm{atm}} \sim 0.05$ eV where $\Delta
m^2_\mathrm{atm}$ is the atmospheric mass-squared difference.
Then, if we adopt as a reasonable guess 
$m_D \sim m_{\mu,\tau}$, the right-handed scale is typically in the
range $m_R \sim 10^8 \div 10^{11}$ GeV. One could also use 
$m_D \sim$ electroweak scale, then $m_R \sim 10^{15}$ GeV could be
identified with the GUT scale.

For the rest of this report we reduce the three sources of neutrino
mixing to one, namely to $M_R$. This means that we choose
\emph{diagonal} coupling matrices $\Gamma_j$ and $\Delta_j$. This is
a well-defined renormalizable theory: diagonal Yukawa couplings
are guaranteed by conservation of the family lepton numbers
$L_\alpha$ ($\alpha = e,\mu,\tau$) which are \emph{softly} broken by
the Majorana mass term of the $\nu_R$
in the Lagrangian of Eq.~(\ref{Lsee}) \cite{GL01,GLnondec}. 
We may summarize the properties of such a theory of the seesaw
mechanism in the following way:
\begin{itemize}
\item[$\ast$]
Soft $L_\alpha$ breaking by the $\nu_R$ mass terms occurs
at the \emph{high} scale $m_R$;
\item[$\ast$]
With diagonal Yukawa couplings, the matrices 
$M_\ell$, $M_D$ are diagonal as well; 
\item[$\ast$]
$M_R$ is the only source of neutrino mixing; 
\item[$\ast$]
For $n_H > 1$, in the limit $m_R \to \infty$, 
there is a non-decoupling in the scalar sector, stemming
from the neutral-scalar--charged-lepton vertices, 
in the following sense \cite{GLnondec}: 
\begin{itemize}
\item
Amplitudes of, e.g., $\mu \to e \gamma$ and $Z \to e^- \mu^+$ scale with
$1/m_R^2$ for large $m_R$; 
\item
The amplitude of, e.g.,
$\mu \to 3e$ approaches a constant in that limit and is not suppressed
by $m_R$; rather, it is suppressed by 
a product of four Yukawa couplings and, possibly, the branching ratio of this
process is within future experimental reach.
\end{itemize}
\end{itemize}
The models developed in Refs.~\cite{GL01,GL03,GLcp}, which will be
reviewed in the following, are all of this type.

\section{Models for obtaining mass matrix M1}

\subsection{The $\mathbbm{Z}_2$ model}

According to the previous section, the $\mathbbm{Z}_2$ model
of Ref.~\cite{GL01} contains the
SM multiplets supplemented by three right-handed heavy neutrino
singlets $\nu_R$; moreover, it has three Higgs doublets $\phi_j$. 
The symmetries are the following: 
\begin{itemize}
\item
$U(1)_{L_\alpha}$ ($\alpha = e,\mu,\tau$) associated with the family
  lepton numbers $L_\alpha$;
\item
\refstepcounter{equation}
$\label{Z2tr}
\mathbbm{Z}_2^{(\mathrm{tr})}: \;\,\quad 
D_{\mu L} \leftrightarrow D_{\tau L} \, , \: 
\mu_R \leftrightarrow \tau_R\, ,
\nu_{\mu R} \leftrightarrow \nu_{\tau R}\, , \: \phi_3 \to - \phi_3 \,;$ 
\hfill (\arabic{equation})
\item
\refstepcounter{equation}
$\label{Z2aux}
\mathbbm{Z}_2^{(\mathrm{aux})}: \quad
\nu_{eR},\: \nu_{\mu R},\: 
\nu_{\tau R},\: \phi_1,\:
e_R\, \: \mbox{change sign} \,.$
\hfill (\arabic{equation})
\end{itemize}
The symmetry $\mathbbm{Z}_2^{(\mathrm{tr})}$
transposes the muon and tau family and is spontaneously broken by the
VEV of $\phi_3$. The symmetry $\mathbbm{Z}_2^{(\mathrm{aux})}$,
spontaneously broken by the VEV of $\phi_1$, is an 
auxiliary $\mathbbm{Z}_2$ which
prevents---at the tree level---$\mathbbm{Z}_2^{(\mathrm{tr})}$
breaking in the neutrino sector.
The above symmetries determine the Yukawa Lagrangian as 
\begin{equation}\label{L}
\begin{array}{rcl}
\mathcal{L}_\mathrm{Y} & = & 
- y_1 \bar D_{eL} \nu_{eR} \tilde\phi_1  
- y_2 \left( \bar D_{\mu L} \nu_{\mu R} + \bar D_{\tau L} \nu_{\tau R} \right)
\tilde\phi_1 
\\ && 
- y_3 \bar D_{eL} e_R \phi_1
- y_4 \left( \bar D_{\mu L} \mu_R + \bar D_{\tau L} \tau_R \right) \phi_2
\\ &&
- y_5 \left( \bar D_{\mu L} \mu_R - \bar D_{\tau L} \tau_R \right) \phi_3
+ \mbox{H.c.}
\end{array}
\end{equation}

The mass matrix $M_R$ is $S$-invariant, i.e., 
$S M_R S = M_R$. Moreover, 
from Eq.~(\ref{L}) we read off that 
$M_D = \mbox{diag}\, (c,d,d)$, i.e., $M_D$ is $S$-invariant as
well. Consequently, $\mathcal{M}_\nu$ given by Eq.~(\ref{Mseesaw}) is
$S$-invariant and due to Eq.~(\ref{M1def}) has the form M1.

The family lepton number groups $U(1)_{L_\mu}$ and $U(1)_{L_\tau}$
do not commute with $\mathbbm{Z}_2^{(\mathrm{tr})}$; thus the basic
non-abelian symmetry group in the $\mu$--$\tau$ sector is $O(2)$
\cite{su5}. Furthermore, 
it was shown that the $\mathbbm{Z}_2$ model can be
embedded in an $SU(5)$ Grand Unified Theory \cite{su5}.

\subsection{The $D_4$ model}

This model \cite{GL03} has the same multiplets as the $\mathbbm{Z}_2$
model, but we add two real scalar gauge singlets
$\chi_1$ and $\chi_2$. The
symmetries of the $D_4$ model are the following:
\begin{itemize}
\item $\mathbbm{Z}_2^{(\mathrm{tr})}: \; \ldots,\ 
\chi_1 \leftrightarrow \chi_2\,,$ where the dots indicate the
transformations of Eq.~(\ref{Z2tr});
\item
\refstepcounter{equation}
$\mathbbm{Z}_2^{(\tau)}: \;
D_{\tau L},\ \tau_R,\ \nu_{\tau R},\
\chi_2\ \mbox{change sign} \,;$
\hfill (\arabic{equation})
\item
$\mathbbm{Z}_2^{(\mathrm{aux})}$ as in Eq.~(\ref{Z2aux}).
\end{itemize}
The symmetries $\mathbbm{Z}_2^{(\mathrm{tr})}$ and 
$\mathbbm{Z}_2^{(\tau)}$ generate the 2-dimensional irreducible
representation $\underline{2}$ of the group $D_4$. Thus the pairs 
$(D_{\mu L}, D_{\tau L})$, $(\mu_R, \tau_R)$, $(\nu_{\mu R}, \nu_{\tau R})$, 
and $(\chi_1, \chi_2)$ transform all as $\underline{2}$ under $D_4$. 
With the above symmetries we obtain the Yukawa Lagrangian 
\begin{equation}\label{L'}
\mathcal{L}'_\mathrm{Y} = \mathcal{L}_\mathrm{Y} +
\left[ \frac{1}{2}\, y_\chi\, \nu_{eR}^T C^{-1} 
\left( \nu_{\mu R} \chi_1 + \nu_{\tau R} \chi_2 \right)
+ \mbox{H.c.} \right],
\end{equation}
where $\mathcal{L}_\mathrm{Y}$ is given by Eq.~(\ref{L}). 
Furthermore, there is a Majorana mass term of the right-handed neutrino
singlets 
\begin{equation}\label{LM}
\mathcal{L}_\mathrm{M} = \frac{1}{2} \left[
M^\ast \nu_{eR}^T C^{-1} \nu_{eR} + 
{M^\prime}^\ast \left( \nu_{\mu R}^T C^{-1} \nu_{\mu R}
+ \nu_{\tau R}^T C^{-1} \nu_{\tau R} \right) \right] + \mbox{H.c.}
\end{equation}

The mass matrix $M_R$ has not only contributions from
$\mathcal{L}_\mathrm{M}$, but also from the 
VEVs of the $\chi_i$, which may be parameterized in the following way: 
\begin{equation}\label{VEVchi}
\left\langle \chi_1 \right\rangle_0 = W \cos{\gamma}
\,, \quad 
\left\langle \chi_2 \right\rangle_0 = W \sin{\gamma}
\,,
\end{equation}
with $W>0$.
The VEVs of the Higgs doublets represent the electroweak scale via 
$v^2 \equiv \sum_j |v_j|^2 = (246\: \mbox{GeV})^2$. 
According to the seesaw mechanism we assume 
\begin{equation}\label{scales}
W \sim |M| \,, \: |M'| \gg v \,.
\end{equation}
Then, by considering the scalar potential,
one can show \cite{GL03} that $\cos 2\gamma = \mathcal{O}(v^2/W^2)$ or 
$\gamma = 45^\circ$ up to 
corrections of order $v^2/W^2$; such corrections are completely
negligible and, therefore, 
$\left\langle \chi_1 \right\rangle_0 = 
 \left\langle \chi_2 \right\rangle_0 = W/\sqrt{2}$. 
These VEVs break $D_4$ down to $\mathbbm{Z}_2^{(tr)}$.

Finally, we arrive at 
\begin{equation}
M_D = \mbox{diag}\, (c,d,d) \,, \quad 
M_R = \left(
\begin{array}{ccc}
M & M_\chi & M_\chi \\
M_\chi & M' & 0 \\
M_\chi & 0 & M'
\end{array}
\right),
\end{equation}
with 
$M_\chi = y_\chi^\ast W/\sqrt{2}$. 
As in the previous section, 
$M_D$ and $M_R$ are both $S$-invariant and, therefore,
$\mathcal{M}_\nu$ is of the form M1. 

In the $D_4$ model, the effective mass probed in neutrinoless
$\beta\beta$-decay can be expressed by the masses of the light
neutrinos, namely 
$\left| \left\langle m \right\rangle \right| = m_1 m_2/m_3$. This is a
consequence of $\left( M_R \right)_{\mu\tau} = 0$. For a further
discussion of $\left| \left\langle m \right\rangle \right|$ see
Ref.~\cite{GL03}.

\section{Models for obtaining mass matrix M2}

\subsection{$A_4$ models}

Because of its irreducible representations, 
the group $A_4$ of the even permutations of four objects is an interesting
discrete group for model building \cite{rajasekaran}. 
Originally, the mass matrix M2 of Eq.~(\ref{M2}) was obtained in a
supersymmetrized version of the SM 
with additional fermionic and scalar singlets
\cite{ma1}. Then a model without supersymmetry,
where the SM was enlarged by an $A_4$-triplet of charged scalar
singlets of Zee type and heavy gauge singlets $E_{L,R}$, was devised
in Ref.~\cite{ma2}. However, we will not pursue this line but 
concentrate instead on relation (\ref{M2def}) which suggests the use of a
non-standard $CP$ transformation (for a review of the theory of $CP$
transformations see Ref.~\cite{rebelo}).

\subsection{The $CP$ model}

In this model \cite{GLcp} the multiplets are the same as in the
$\mathbbm{Z}_2$ model. The symmetries are the following: 
\begin{itemize}
\item
$U(1)_{L_\alpha}$ ($\alpha = e,\mu,\tau$);
\item
The non-standard $CP$ transformation \cite{GLcp,scott} 
\begin{equation}\label{CP}
\begin{array}{l}
D_{\alpha L} \to i S_{\alpha \beta}
\gamma^0 C \bar D_{\beta L}^T\, , \quad  
\nu_{\alpha R} \to i S_{\alpha \beta}
\gamma^0 C \bar \nu_{\beta R}^T\, , \quad
\alpha_R \to i S_{\alpha \beta}
\gamma^0 C \bar \beta_R^T\, , \\ 
\phi_{1,2} \to \phi_{1,2}^\ast\, , \quad
\phi_3 \to - \phi_3^\ast \,,
\end{array}
\end{equation}
where $\alpha, \beta = e, \mu, \tau$ and $S$ is defined in Eq.~(\ref{S});
\item
$\mathbbm{Z}_2^{(\mathrm{aux})}$.
\end{itemize}
With these symmetry operations, we obtain the Yukawa Lagrangian 
\begin{equation}
\begin{array}{rcl}
\mathcal{L}''_\mathrm{Y} & = & 
- y_1 \bar D_e \nu_{eR} \tilde\phi_1  
- \left( y_2 \bar D_\mu \nu_{\mu R} + y_2^* \bar D_\tau \nu_{\tau R} \right)
\tilde\phi_1 
\\ && 
- y_3 \bar D_e e_R \phi_1
- \left( y_4 \bar D_\mu \mu_R +  y_4^* \bar D_\tau \tau_R \right) \phi_2
\\ &&
- \left( y_5 \bar D_\mu \mu_R -  y_5^* \bar D_\tau \tau_R \right)
\phi_3 
+ \mbox{H.c.} 
\end{array}
\label{L''}
\end{equation}
The coupling constants $y_1$ and $y_3$ are real,
whereas $y_2$, $y_4$, and $y_5$ are in general complex. 

Assuming without loss of generality
$v_1 \in \mathbbm{R}$, we have 
$M_D = \mbox{diag}\, (c, d, d^*)$
with $c \in \mathbbm{R}$, and therefore $M_D$ fulfills
$M_D^\ast = S M_D S$. Since, by virtue of
Eq.~(\ref{CP}), $M_R^\ast = S M_R S$ holds, it follows that 
$\mathcal{M}_\nu$ fulfills relation (\ref{M2def}) and has, therefore,
the form M2.
We note that in the $CP$ model $m_\mu \neq m_\tau$ is a
consequence $CP$ violation \cite{GLcp}. 

\section{Summary}

In this report we have first discussed the phenomenology of the neutrino mass
matrices M1 and M2 of Eqs.~(\ref{M1}) and (\ref{M2}), respectively,
Then we have reviewed the $\mathbbm{Z}_2$ model of Ref.~\cite{GL01} and
the $D_4$ model of Ref.~\cite{GL03}, which both yield the mass matrix
M1, and the $CP$ model of Ref.~\cite{GLcp}, which yields mass matrix
M2. These three models have several features in common: 
\begin{itemize}
\item[$\Join$]
The SM is enlarged with three right-handed neutrino singlets, there
are three Higgs doublets instead of one, and---only in the case of
the $D_4$ model---there are two real scalar gauge singlets. 
\item[$\Join$]
The seesaw mechanism is responsible for the smallness of the neutrino
masses. 
\item[$\Join$]
Below the seesaw scale, the family lepton numbers $L_\alpha$ are softly
broken by the mass term of the right-handed singlets, i.e., 
$M_R$ is the sole source of neutrino mixing.\footnote{We want to
stress that this means also that the charged-lepton mass matrix is
diagonal not by assumption but by virtue of the lepton numbers $L_\alpha$.}
\item[$\Join$]
Neutrino mass matrices of form M1, M2 are obtained by non-abelian
horizontal symmetry groups in the case of the $\mathbbm{Z}_2$ and
$D_4$ models, and by a non-standard $CP$ transformation which does not
commute with $U(1)_{L_\alpha}$ ($\alpha = \mu, \tau$) in the case of
the $CP$ model. 
\item[$\Join$]
All three models have a
maximal atmospheric neutrino mixing  $\theta_{23} = 45^\circ$. 
\item[$\Join$]
The models have no predictions for the neutrino mass spectrum, i.e.,
there are no relations among the masses or between the masses and
mixing angles. 
\end{itemize}
Other predictions for the mixing angles are 
$\sin \theta_{13} = 0$ in the case of matrix
M1, and $\sin \theta_{13} \neq 0$, $e^{i\delta} = \pm i$ in the case
of matrix M2. 
Finally we note that, looking at the Yukawa Lagrangians of Eqs.~(\ref{L}),
(\ref{L'}), and (\ref{L''}), one would expect the ``natural relation''
$m_\mu \sim m_\tau$, following from the $\mu$--$\tau$ interchange symmetry.
However, simply by introducing an additional
symmetry but no further multiplets one can achieve $m_\mu \ll m_\tau$
in a technically natural way \cite{mutau}.

\vspace{4mm}

\noindent
\textbf{\Large Acknowledgment:}\\ 
W.G. would like to thank the organizers for their great hospitality.

\end{document}